\documentclass[11pt]{article}
\usepackage{amsfonts}
\usepackage{graphicx}
\usepackage{epstopdf}
\usepackage{epsfig}
\usepackage{amssymb}
\usepackage{setspace}
\usepackage{caption}
\usepackage{amsfonts}
\usepackage{color}
\usepackage{amsmath}
\usepackage{float}
\usepackage{comment}
\newcommand {\be}{\begin{equation}}
\newcommand {\ee}{\end{equation}}
 \newcommand {\bea}{\begin{array}}
 
 \newcommand {\eea}{\end{array}}
\newcommand{\RN}{Reissner-Nordstrom~}

\textwidth=6.5in \hoffset=-.75in \voffset=-1in \numberwithin{equation}{section}
\numberwithin{figure}{section}

\begin{document}

\begin{titlepage}
\vspace{1cm} 
\begin{center}
{\Large \bf {Rotating and charged Taub-NUT-(A)dS spacetimes on a 3-brane}}\\
\end{center}
\vspace{2cm}
\begin{center}
	\renewcommand{\thefootnote}{\fnsymbol{footnote}}
	Haryanto M. Siahaan{\footnote{haryanto.siahaan@unpar.ac.id}}\\
	Center for Theoretical Physics, Department of Physics,\\
	Parahyangan Catholic University,\\
	Jalan Ciumbuleuit 94, Bandung 40141, Indonesia
	\renewcommand{\thefootnote}{\arabic{footnote}}
\end{center}

\begin{abstract}

We construct novel solutions to the effective Einstein equation with four dimensional cosmological constant on a 3-brane in Randall-Sundrum II scenario. The charged solution is obtained by assuming the existence of localized Maxwell fields on the 3-brane. Timelike and null circular geodesics in the neutral rotating braneworld spacetime with NUT parameter are discussed, and we find that such geodesics cannot occur on the equatorial plane. On non-equatorial planes, timelike circular geodesics can occur for the negative tidal charge, but we cannot find numerical results to support the existence of non-equatorial null circular geodesics.

\end{abstract}
\end{titlepage}\onecolumn 
\bigskip 

\section{Introduction}
\label{sec:intro}

String theory or M-theory \cite{Kiritsis:1997hj,Townsend:1996xj,Becker} predicts the existence of extra dimensions in addition to our daily experienced four dimensional spacetime. The idea of extra dimensions goes back to the attempt by Kaluza \cite{Kaluza:1921tu} and Klein \cite{Klein:1926tv} in unifying gravity and electromagnetism by considering five dimensional Einstein theory where the extra spatial dimension is compactified into a circle $S^1$ with very small radius. The coordinate transformation of this compact manifold $S^1$ then associates the Abelian gauge invariance of Maxwell theory, and leaving the residual spacetime dimensions to have general coordinate invariance of the ordinary Einstein gravitational theory. 

The idea that our four dimensional world could be a hypersurface of a higher dimensional spacetime, which we would refer as braneworld scenario, had appeared almost four decades ago in the work by Rubakov et al. \cite{Rubakov:1983bb}. The four dimensional spacetime where we live is normally referred as the brane, embedded in the higher dimensional spacetime which is called as the bulk. In the braneworld model of Rubakov et al., the geometry under consideration is Minkowski space $M^{\left(3+N,1\right)}$ with $\left(3+N\right)$ spatial dimensions. Clearly this type of braneworld picture has not incorporated gravity. Light or ordinary particles are confined inside a potential well that is narrow enough along the $N$ spatial dimensions. Almost two decades later, the quite similar idea was used to tackle the hierarchy problem \cite{ArkaniHamed:1998nn,ArkaniHamed:1998rs,Randall:1999ee,Randall:1999vf,Dvali:2000hr,Karch,Maartens:2010ar,Tanahashi:2011xx}, where the braneworld models include gravity into consideration. In these models, unlike the matter fields which are assumed to be localized on the 3-brane\footnote{From now on, we refer the 3-brane just as the brane.}, gravity can propagate in extra dimensions. 

Since gravity is an integral part in the braneworld approach, black holes must be among the main interests. Particularly in the Randall-Sundrum II (RS-II) model \cite{Randall:1999vf} which we consider in this work, Shiromizu et al. \cite{Shiromizu} have provided us the effective Einstein equation on the brane using Gauss-Codazzi approach. This equation has been studied extensively, and the corresponding braneworld black hole solutions have been proposed. For example the static black hole solution to this effective equation was given in \cite{Dadhich}, where the metric function has the form of \RN solution with a ``charge'' term appeared in the metric. This charge is interpreted as the tidal charge and considered as the bulk geometrical effect. If we assume that the brane is not vacuum, for example it contains localized Maxwell fields, this tidal charged has no coupling to the $U\left(1\right)$ gauge fields. This implies there is no electromagnetic interaction between charged probe and the neutral braneworld black hole with tidal charge reported in\cite{Dadhich}. The non-vacuum brane in RS-II model containing localized Maxwell fields was considered in \cite{Chamblin}, where the static charged black hole analogous to the \RN solution of Einstein-Maxwell theory was given. The limit of vanishing electric charge in the solution by Chamblin et al. \cite{Chamblin} reduces to that in \cite{Dadhich}, and these black holes are asymptotically flat. The rotating and charged solution of black holes in RS-II model was obtained by Aliev et al. \cite{Aliev:2005bi}, using the Kerr-Schild ansatz in solving the Hamiltonian constraint from the effective Einstein equation on the brane. The case of (A)dS black holes on the brane were investigated in \cite{Neves:2012it}, where the effective Einstein equation under consideration contains the four dimensional cosmological constant. Furthermore, topological black holes on the brane were studied in \cite{Sheykhi:2008et} where a class of solution describes the static charged AdS black hole on the brane. Braneworld black holes from the bulk black string solutions in five-dimensional theory with a scalar field
non-minimally coupled to gravity are studied in \cite{Kanti:2018ozd,Nakas:2019rod,Nakas:2020crd}.

Assuming the existence of localized black holes on the brane, black holes or compact objects in braneworld scenario have been discussed in several aspects. Gravitational wave which is produced by collisions of massive objects in braneworld scenario could open the window to search for extra dimensions \cite{Visinelli:2017bny}. Studies related to the lensing and optical process for braneworld black holes are presented in \cite{BinNun:2009jr,Whisker:2004gq,Eiroa:2012fb,Abdujabbarov:2017pfw,Eiroa:2004gh}, and the possibility to constrain the extra dimension parameter namely the bulk AdS$_5$ radius from the black hole image found by The Event Horizon Telescope assuming that the supermassive rotating black hole at the center of M87 galaxy is described by RS-II braneworld black hole \cite{Vagnozzi:2019apd}. However, despite the number of studies for localized black holes as previously mentioned, the localization of black hole on the brane itself is still under investigation \cite{Fitzpatrick:2006cd,Tanaka:2002rb,Emparan:2002px,Kanti:2001cj,Kanti:2003uv,Kanti:2013lca,Kanti:2015poa,Fichet:2019owx}. The first numerical study of the gravitational collapse of a strong
pulse of massless scalar field within the framework of the RS-II mode is a black hole with finite extension into the bulk \cite{Wang:2016nqi}.

The Kerr-Newman-Taub-NUT-(A)dS spacetime is a class of solutions to the Einstein-Maxwell theory with a non vanishing cosmological constant $\Lambda$. This solution is a special case of Plebanski-Demianski solution \cite{Plebanski:1976gy,Griffiths:2009dfa} that has parameters of mass, electric charge, rotation, NUT, and cosmological constant. With or without the electric charge and cosmological constant in the discussions, this solution has been studied rigorously in literature \cite{Houri:2019lnu,Kolar:2019gzy,Chakraborty:2019rna,Long:2018tij,Frolov:2018eza,Sadeghian:2018bli,Mukherjee:2018dmm,Krtous:2018bvk,Duztas:2017lxk,Paganini:2017qfo}. Despite the irregularity of spacetime with NUT parameter on its rotational axis \cite{Cebeci:2017xex,Cebeci:2015fie,Siahaan:2019kbw}, most aspects of this type of spacetime resemble those of black holes \cite{Sakti:2017pmt,Sakti:2019krw,Sakti:2019zix,Sakti:2020jpo,Jefremov:2016dpi}. The NUT parameter itself has an interesting interpretation\cite{Griffiths:2009dfa}, i.e. the gravitomagnetic mass, after making an analogy to the electromagnetic phenomena where gravitational mass in the Kerr-Taub-NUT solution sometime is labeled as the gravitoelectric charge. Furthermore, the presence of NUT parameter in Kerr-Taub-NUT solution makes the asymptotic geometry to be non-flat, unlike the Kerr metric counterpart. On the other hand, the presence of NUT parameter yields the absence of curvature singularity in the spacetime, despite the conic singularity that it has. This conic singularity problem somehow can be solved by introducing periodicity in timelike coordinate, but it costs the existence of closed timelike curves in spacetime. Consequently, the collapsing object in Kerr-Taub-NUT spacetime or its generalization cannot be considered as the regular black hole. Nevertheless, investigating properties of rotating object with NUT parameter is still of great interests, for example the motions of test objects around it. It was shown that equatorial circular geodesics cannot exist in rotating spacetime with NUT parameter, in Einstein-Maxwell theory or low energy heterotic string version \cite{Jefremov:2016dpi,Siahaan:2019kbw}.

In this paper, we extend the rotating charged black hole solution proposed by Aliev and G\"umr\"uk\c{c}\"uo\~glu \cite{Aliev:2005bi}, where now the solution includes NUT parameter and cosmological constant $\Lambda$. We follow the same prescription presented in \cite{Aliev:2005bi}, where the spacetime ansatz takes the form of Kerr-Schild, and solve the corresponding equation which is the trace of effective equation on the brane in RS-II scenario. The class solution introduced in this work resembles the Kerr-Newman-Taub-NUT-(A)dS solution of Einstein-Maxwell theory, with additional tidal charge as an effect from the bulk. Particularly for the spacetime of rotating neutral black hole with NUT parameter, we investigate the corresponding equatorial circular geodesics by deriving the associated Hamilton-Jacobi equation first. The organization in this paper is as follows. In the next section, we give a short review on the effective gravitational equations on brane whose solutions we attempt to achieve in this work.  In section \ref{sec.sol3brane}, several types solutions are constructed. We start by getting the neutral rotating spacetime with tidal charge and NUT parameter in section \ref{ssec.KerrTaubNUTon3brane}, followed by including cosmological constant in section \ref{ssec.addingcosmo}. The rotating charged case is discussed in section \ref{ssec.incorporatingcharge}, while the non-rotating limit is considered in section \ref{ssec.staticchargedAdS}. The investigation of equatorial circular geodesics around rotating object on the brane with tidal charge and NUT parameter is performed in section \ref{sec.geodesicsaroundblackhole}. Finally we give conclusion and discussion. In this paper, we consider $c=G_4=1$ where $c$ and $G_4$ are the light speed and four dimensional cosmological constant, respectively. 

\section{Effective equations on the brane}\label{sec.effectiveeqtn3brane}

In their seminal work \cite{Shiromizu}, Shiromizu et al. obtained an effective Einstein equation on the brane in the context of RS-II scenario\cite{Randall:1999vf}. The Einstein equation in the bulk reads
\be\label{Einstein.eqtn.bulk}
^{\left( 5 \right)} G_{MN}  + \Lambda _5 \bar g_{MN}  = \kappa _5^2 \left( {^{\left( 5 \right)} T_{MN}  + \sqrt {\frac{g}{{\bar g}}} {\cal T}_{MN} \delta \left( X \right)} \right)
\ee 
with $\kappa _5^2 = 8\pi G_5$, and $G_5$ is the five dimensional Newton gravitational constant. Intuitively $^{\left( 5 \right)} G_{MN}$, $^{\left( 5 \right)} T_{MN}$, and $\Lambda_5$ are five dimensional Einstein tensor, energy-momentum tensor, and bulk cosmological constant, respectively. Bulk spacetime metric is denoted by ${\bar g}_{MN}$, while the brane is endowed with the metric $g_{MN}$. Then $\bar g$ and $g$ represent the determinant of bulk and brane spacetime metric, respectively. If we allow the brane to be non-empty, then the energy and momentum tensor distribution on the brane is represented by ${\cal T}_{MN}$. The scalar function $X=X\left(y^M\right)$ is introduced to specify the ``brane'' hypersurface, namely $X={\rm const.}$, where the hypersurfaces $\Sigma_X$ have no timelike intersections between each other. The coordinates $y^M$ labels the spacetime in the bulk
with the capital Latin indices $M=0,1,2,3,4$.

Deriving the effective Einstein equation on the brane can be done by employing the Gauss-Codazzi method to project the bulk equation onto the brane, and also imposing ${\mathbb Z}_2$ symmetry to get the desired equation. In \cite{Shiromizu}, the choice of Gaussian normal coordinate were subsequently picked in particular form. However, a more general setting for the coordinates to get the effective Einstein equation on the brane from the five dimensional gravitational in the bulk \cite{Aliev:2004ds}. The effective Einstein equations on the brane with cosmological constant can be shown to take the form \cite{Shiromizu,Aliev:2005bi}
\be \label{eq.EinsteinON3brane}
G_{\mu \nu } + \Lambda_{4} g_{\mu\nu}  = 8\pi T_{\mu \nu }  + \kappa _5^4 \left( {S_{\mu \nu }  - \frac{1}{2}g_{\mu \nu } S_\alpha ^\alpha  } \right) - E_{\mu \nu } \,,
\ee
where $G_{\mu \nu }$ is the four dimensional Einstein tensor, $\Lambda_4$ denotes the four dimensional brane cosmological constant, and $g_{\mu \nu }$ is the corresponding tensor metric. In the next section we will expressed the five dimensional cosmological constant in terms of AdS$_5$ curvature radius $\ell$ as
\be \label{Lambda5}
\Lambda _5  =  - \frac{6}{{\ell^2 }}\,,
\ee 
and relation between five and four dimensional Newton constants is $G_5 = \ell G_4$. In this paper we use convention $G_4 =1$ which then implies $G_5 =\ell$ and $\kappa _5^4 = 64 \pi^2 \ell^2$. Explicitly the AdS radius $\ell = 6\left(\lambda \kappa_5^2\right)^{-1}$ where $\lambda$ denotes the brane tension. 

The bulk contribution in eq. (\ref{eq.EinsteinON3brane}) is given by the traceless tensor
\be \label{eq.Eij}
E_{\mu \nu }  = ^{\left( 5 \right)} C_{KLMN} n^K n^M e_\mu ^L e_\nu ^N 
\ee 
which comes from the projection of five dimensional Weyl tensor on the brane. In eq. (\ref{eq.Eij}) we have used the unit spacelike vector $n^K$ normal to the brane and tangent vectors representing local frame
\be \label{legs}
e_\mu ^K  = \frac{{\partial y^K }}{{\partial x^\mu  }}\,
\ee 
obeying $n_K e_\nu ^K  = 0$. Here we have used $x^\mu$ as the brane spacetime coordinate where $\mu=0,1,2,3$. Using the legs in (\ref{legs}), the bulk and brane spacetime metric can be connected via
\be
\bar g_{MN}  = n_M n_M  + g_{\mu \nu } e_M^\mu  e_N^\nu  \,.
\ee 
As it was argued in \cite{Dadhich,Chamblin,Aliev:2005bi,Neves:2012it}, the traceless tensor $E_{\mu\nu}$ is assumed to exist in order to fulfill the corresponding Einstein equations. This tensor comes from the bulk geometrical effect, and until today the exact bulk spacetime geometry which gives the exact brane spacetime as reported in \cite{Dadhich,Chamblin,Aliev:2005bi,Neves:2012it} have not been discovered yet. In the next sections, we follow the same spirit, assuming the corresponding $E_{\mu\nu}$ to exist, and the spacetime metric component can be achieved using the trace of effective Einstein equation. 

In this work, we will consider the brane to be filled by some localized Maxwell fields. This field gives rise to the electric charge of a black hole, and electromagnetic interaction between the black hole and charged object. Therefore, the associated energy-momentum tensor on the brane reads 
\be\label{Tmn}
T_{\mu \nu }  = \frac{1}{{4\pi }}\left( {F_{\mu \alpha } F_\nu ^\alpha   - \frac{1}{4}g_{\mu \nu } F_{\alpha \beta } F^{\alpha \beta } } \right)\,,
\ee 
and the ``squared'' of $T_{\mu\nu}$ in (\ref{eq.EinsteinON3brane}) is given by
\be
S_{\mu \nu }  =  - \frac{1}{4}\left( {T_{\mu \alpha } T_\nu ^\alpha   - \frac{1}{2}g_{\mu \nu } T_{\alpha \beta } T^{\alpha \beta } } \right)\,.
\ee
The field strength tensor takes the familiar form, namely $F_{\mu\nu} = \partial_\mu A_\nu - \partial_\nu A_\mu$. Nevertheless, the absence of an effective action which is responsible for the equation of motion in (\ref{eq.EinsteinON3brane}) hinders us to have the second field equation, in addition to (\ref{eq.EinsteinON3brane}), which is supposedly obtained by varying an action with respect to $A_\mu$. However, we can assume that the source free equation $\nabla_{\mu} F^{\mu\nu}=0$ holds when we discuss the brane with confined electromagnetic fields.

\section{Solutions on the brane}\label{sec.sol3brane}

\subsection{Kerr-Taub-NUT spacetime with tidal charge on the brane}\label{ssec.KerrTaubNUTon3brane}

The neutral rotating braneworld black hole in RS-II scenario was found in \cite{Aliev:2005bi} has the spacetime metric that can be expressed as 
\be \label{metric.BWKerr}
 {\rm{d}}s^2  =  - {\frac{{\Delta_0}}{{r^2  + a^2 x^2 }}} \left( {{\rm{d}}t - a\Delta _x {\rm{d}}\phi } \right)^2  + \left(r^2+a^2 x^2\right)\left(\frac{{\rm d}r^2}{\Delta_0} + \frac{{\rm d}x^2}{\Delta_x}\right) + \frac{{\Delta _x \left( {a{\rm{d}}t - \left( {r^2  + a^2 } \right){\rm{d}}\phi } \right)^2 }}{{\left( {r^2  + a^2 x^2 } \right)}}\,,
\ee
where we have used $\Delta_x= 1-x^2$ and $\Delta_0 = r^2 - 2Mr +a^2 +\beta$. This metric is just the Kerr-Newman solution after replacing $\beta\to Q^2$ where $Q$ is charge of the black hole. Here, $\beta$ is known as the tidal charge \cite{Dadhich,Aliev:2005bi} and considered as some geometrical effects from the bulk. This spacetime metric is asymptotically flat, and possesses the timelike and axial Killing symmetries. Using Komar integrals, these Killing symmetries can lead to two conserved charges, namely the black hole mass $m$ and angular momentum $J = aM$. 

In this section, we consider the simplest extension to the solution (\ref{metric.BWKerr}) where now the spacetime is equipped with NUT parameter $n$. Recall that the Kerr-Taub-NUT spacetime is solution to the vacuum Einstein equation, hence we can keep considering the vacuum brane as in the case described by the metric (\ref{metric.BWKerr}). To proceed, we follow the prescription by Aliev et al. \cite{Aliev:2005bi} using the Kerr-Schild ansatz for the metric and solve the Hamiltonian constraint equation for the unknown function in the metric. Since we are interested in the metric solution with NUT parameter, the ``massless'' term in Kerr-Schild ansatz is extended to contain the NUT parameter, which then yields the Kerr-Schild ansatz to be
\be \label{metric.mNUTH}
{\rm d}s^2  = {\rm d}\bar s_{{\rm{mNUT}}}^{\rm{2}}  + H\left( {r,x} \right)\left( {l_\mu  {\rm d}x^\mu  } \right)^2 \,,
\ee 
where\footnote{We use mNUT acronym for massless-Newman-Unti-Tamburino.}
\[
{\rm d}\bar s_{{\rm{mNUT}}}^{\rm{2}}  =  - \frac{{r^2  - n^2  + a^2 x^2 }}{{\rho ^2 }}{\rm d}u^2  + \frac{{\Delta _x \left( {r^2  + a^2  + n^2 } \right)^2  - P^2 \left( {r^2  + a^2  - n^2 } \right)}}{{\rho ^2 }}{\rm d}\psi ^2  + \frac{{\rho ^2 }}{{\Delta _x }}{\rm d}x^2 
\]
\be \label{KSansatz-1}
- \frac{{4n\left( {\left( {r^2  + a^2  - n^2 } \right)x + an\Delta _x } \right)}}{{\rho ^2 }}{\rm d}\psi du + 2{\rm d}u{\rm d}r + \left( {4nx - 2a\Delta _x } \right){\rm d}\psi {\rm d}r\,.
\ee
Here we have used $\rho^2 = r^2 + \left(n +ax\right)^2$ and the null vector $l_\mu  {\rm d}x^\mu   = {\rm d}u - P{\rm d}\psi $ where $P = a\Delta_x - 2nx$. Taking the limit $n\to 0$ in the ${\rm d}\bar s_{{\rm{mNUT}}}^{\rm{2}}$ gives us the massless limit of Kerr metric. The quantities $n$ and $a$ are the NUT and rotational parameters, respectively. On the brane, from eq. (\ref{eq.EinsteinON3brane}) one can write the vacuum effective Einstein equation without the four dimensional cosmological constant $\Lambda_4$ as
\be  \label{eq.EinsteinVACUUMbrane}
R_{\mu\nu}  =  - E_{\mu\nu}\,.
\ee 
As the result that $E_{\mu\nu}$ being tracefree, the corresponding Hamiltonian constraint achieved from the last equation is
\be \label{eq.Einstein3braneVACUUM}
R = 0 \,.
\ee 

Plugging the metric ansatz (\ref{KSansatz-1}) into eq. (\ref{eq.Einstein3braneVACUUM}) gives us
\be \label{eqH}
\left( {\frac{{\partial ^2 }}{{\partial r^2 }} + \frac{{4r}}{{\rho ^2 }}\frac{\partial }{{\partial r}} + \frac{2}{{\rho ^2 }}} \right)H\left( {r,x} \right) = 0\,.
\ee 
As one would expect, eq. (\ref{eqH}) reduces to the analogous one appearing in \cite{Aliev:2005bi} as the limit $n\to 0$ is taken. Solution to the last equation can be written as
\be\label{H.sol.NUT}
H\left( {r,x} \right) = \frac{{2Mr - \beta }}{{\rho ^2 }}\,,
\ee 
where $m$ and $\beta$ in the last expression are interpreted as the mass and tidal charge parameters of the object on the brane. To achieve the Boyer-Lindquist form of the metric for (\ref{metric.mNUTH}) with the solution $H\left(r,x\right)$ in (\ref{H.sol.NUT}), we can make use the following transformation
\be \label{BLtransmNUT}
{\rm d}u = {\rm d}t + \frac{{r^2  + a^2  + n^2 }}{{\Delta_r }}{\rm d}r
~~{\rm and}~~
{\rm d}\psi  = {\rm d}\phi  + \frac{a}{{\Delta_r }}{\rm d}r\,,
\ee 
with $\Delta _r  = r^2  - 2Mr + a^2  - n^2  + \beta$. Explicitly, the resulting metric after performing transformation (\ref{BLtransmNUT}) to the eq. (\ref{metric.mNUTH}) can be written as
\be \label{metric.KNUT.brane}
{\rm{d}}s^2  =  - \frac{{\Delta _r }}{{\rho ^2 }}\left( {{\rm{d}}t - P{\rm{d}}\phi } \right)^2  + \rho ^2 \left( {\frac{{{\rm{d}}r^2 }}{{\Delta _r }} + \frac{{{\rm{d}}x^2 }}{{\Delta _x }}} \right) + \frac{{\Delta _x }}{{\rho ^2 }}\left( {\left( {r^2  + a^2  + n^2 } \right){\rm{d}}\phi  - a{\rm{d}}t} \right)^2 \,.
\ee 
In the absence of tidal charge $\beta$, the metric (\ref{metric.KNUT.brane}) is just the Kerr-Taub-NUT line element, i.e. solution to the vacuum Einstein equation $R_{\mu\nu} = 0$. Furthermore, the line element (\ref{metric.KNUT.brane}) is just the Kerr-Newman-Taub-NUT metric after replacing $\beta \to Q^2$, where $Q$ is the electric charge in Kerr-Newman-Taub-NUT spacetime. Note that unlike the charge squared $Q^2$ in Kerr-Newman family which always takes positive value, the tidal charge can be negative which could lead to some non-trivial deviations in the geodesics compared to that of Kerr-Newman-Taub-NUT case. Moreover, the negative tidal charge can help the production of naked singularity which is not allowed to exist according to the cosmic censorship conjecture. 

The tensor $E_{\mu\nu}$ that responsible for the solution above can be computed using eq. (\ref{eq.EinsteinVACUUMbrane}). The components are
\[
E_{tt}  =   -\frac{{\beta }}{{\rho ^6 }} \left( {\Delta _r  + a^2\Delta _x } \right)~,~E_{rr}  = -\frac{\Delta_x}{\Delta_r} ~,~E_{xx}= \frac{\beta }{{\rho ^2 \Delta _r }}\,,
\]
\[
E_{t\phi }  =  - \frac{{2\beta }}{{\rho ^6 }} \left( {nx\Delta _r  - a\Delta _x \left( {a^2  + r^2  + \beta /2} \right) + mra\Delta _x } \right)\,,
\]
\be 
E_{\phi \phi }  = \frac{{\beta }}{{\rho ^6 }} \left( {2mr\left( {2nx - a\Delta _x } \right)^2  - \Delta _x \left( {\left( {a^2  + r^2 } \right)\left( {a^2 \Delta _x  + a^2  + r^2 } \right) + a^2 B\Delta _x } \right) - n\sum\limits_{k = 0}^3 {c_k a^k } } \right)\,,
\ee
with each $c_k$'s in equation above is given by
\[
c_0  = n\left( {n^2 \left( {1 - 5x^2 } \right) + 2r^2 \left( {1 + x^2 } \right) + 4\beta x^2 } \right)\,,
\]
\be 
c_1  = 4x\Delta _x \left( {n^2  - r^2  - \beta } \right)~,~c_2  = n\left( {1 + 4x^2  - x^4 } \right)~,~c_3  =  - 4x\Delta _x \,.
\ee
As it was mentioned before that we assume the existence of bulk metric which gives rise to this components of $E_{\mu\nu}$ following mechanism described in previous section. Obviously this tensor $E_{\mu\nu}$ is proportional\footnote{They become equal after replacing $\beta\to Q^2$ in $E_{\mu\nu}$, multiplied by $-8\pi$.} to the energy-momentum tensor that belongs to Kerr-Newman-Taub-NUT solution, whose field strength can be written as
\be \label{FmnQtrxphi}
{\rm \bf F} = \frac{Q}{{\rho ^4 }}\left[ {\left( {r^2  - \left( {n + ax} \right)^2 } \right){\rm{d}}r \wedge \left( {{\rm{d}}t - P{\rm{d}}\phi } \right) - 2r\left(n+ax\right){\rm{d}}x \wedge \left( {\left( {r^2  + a^2  + n^2 } \right){\rm{d}}\phi  - a{\rm{d}}t} \right)} \right]\,.
\ee 

\subsection{Adding four dimensional cosmological constant}\label{ssec.addingcosmo}

Now let us consider a more general case, where now we include the four dimensional cosmological constant $\Lambda_4$ in our effective Einstein equation on the brane. The equation of motion then reads 
\be\label{eq.Einstei3braneLambda}
G_{\mu \nu }  + \Lambda _4 g_{\mu \nu }  =  - E_{\mu \nu } \,.
\ee 
Adopting the same strategy as in the previous section, the proper metric ansatz can be expressed as
\be\label{metric.KerrSchildmNUTLambda}
ds^2  = {\rm{d}}s_{{\rm{mNUT,\Lambda }}}^2  + H\left( {r,x} \right)\left( {l_\mu  {\rm d}x^\mu } \right)^2 \,,
\ee 
where
\[
{\rm{d}}s_{{\rm{mNUT,\Lambda }}}^2 = 2\left( {{\rm{d}}u{\rm{d}}r - P{\rm{d}}r{\rm{d}}\psi } \right) + \frac{{\rho ^2 }}{{\Xi \Delta _x }}{\rm{d}}x^2  + \frac{{\Xi \Delta _x a^2  - L_r }}{{\rho ^2 }}{\rm{d}}u^2 + \frac{{\Xi \Delta _x \left( {r^2  + a^2  + n^2 } \right)^2  - P^2 L_r }}{{\rho ^2 }}d\psi ^2 
\]
\be 
+  \frac{{2\left( {PL_r - \Xi a\Delta _x \left( {r^2  + a^2  + n^2 } \right) } \right)}}{{\rho ^2 }}{\rm{d}}u{\rm{d}}\psi\,.
\ee 
In the equation above, the null vector is still $l_\mu  {\rm d}x^\mu   = {\rm d}u - P{\rm d}\psi $ with $P = a\Delta_x - 2nx$,
\be 
\Xi  = 1 + \frac{{\Lambda ax}}{3}\left( {4n + ax} \right)\,,
\ee
and
\be 
L_r  = r^2  + a^2  - n^2  - \frac{\Lambda }{3}\left( {3n^2 \left( {a^2  - n^2 } \right) + r^2 \left( {a^2  + 6n^2 } \right) + r^4 } \right)\,.
\ee

The corresponding Hamiltonian constraint equation in this case is the trace of (\ref{eq.Einstei3braneLambda}) which takes the form
\be \label{Hamilton.Lambda}
R = 4\Lambda \,.
\ee 
Applying this equation to the metric ansatz (\ref{metric.KerrSchildmNUTLambda}) gives us exactly eq. (\ref{eqH}) whose solution is given in (\ref{H.sol.NUT}). To obtain the Boyer-Lindquist type metric from the line element (\ref{metric.KerrSchildmNUTLambda}) where the corresponding $H\left(r,x\right)$ is given in (\ref{H.sol.NUT}), we can perform the following transformations
\be
{\rm{d}}u = {\rm{d}}t + \frac{{r^2  + a^2  + n^2 }}{{\Delta_{r,\Lambda}}}{\rm{d}}r~~{\rm and}~~{\rm{d}}\psi  = {\rm{d}}\phi  + \frac{a}{{\Delta_{r,\Lambda}}}{\rm{d}}r\,,
\ee
where $\Delta_{r,\Lambda} = L_r -2Mr +\beta$. Then the resulting metric can be expressed as
\be \label{metric.braneKNUTLambda.boyerlind}
{\rm{d}}s^2  =  - \frac{{\tilde \Delta _r }}{{\rho ^2 }}\left( {{\rm{d}}t - P{\rm{d}}\phi } \right)^2  + \rho ^2 \left( {\frac{{{\rm{d}}r^2 }}{{\tilde \Delta _r }} + \frac{{{\rm{d}}x^2 }}{{\Xi \Delta _x }}} \right) + \frac{{\Xi \Delta _x }}{{\rho ^2 }}\left( {a{\rm{d}}t - \left( {r^2  + a^2  + n^2 } \right){\rm{d}}\phi } \right)^2 \,.
\ee
As one would expect, this solution coincide to the Kerr-Newman-Taub-NUT-(A)dS metric\footnote{See appendix \ref{app.KNnutAdS} for some features of Kerr-Newman-Taub-NUT-(A)dS solution.} after replacing $\beta \to Q^2$. Then one can understand the corresponding traceless tensor $E_{\mu\nu}$ in eq. (\ref{eq.Einstei3braneLambda}) is proportional to the energy-momentum tensor in the Kerr-Newman-Taub-NUT-(A)dS solution after replacing $Q^2\to \beta$. As before, we have assumed that there exist some bulk geometries which allow this $E_{\mu\nu}$ tensor on the brane obeying eq. (\ref{eq.Eij}). The Kerr-(A)dS spacetime on the brane reported in \cite{Neves:2012it} can be obtained from eq. (\ref{metric.braneKNUTLambda.boyerlind}) after setting the NUT parameter to be vanished.

\subsection{Incorporating electric charge on the brane}\label{ssec.incorporatingcharge}

In the previous sections we were dealing with an empty brane, hence the effective Einstein equation on the brane does not contain any energy momentum tensor for fields that are localized on the brane. Here let us consider the brane to be non-empty, filled by electromagnetic field confined on the brane. The energy-momentum tensor for this field takes the familiar form (\ref{Tmn}). Recall that we are working in the unit that $G_4 =1$, then we have $G_5 = \ell$ and $\kappa_5^2 = 8\pi \ell$. Therefore, the effective Einstein equation on the brane in the presence of cosmological constant and Maxwell fields in the brane (\ref{eq.EinsteinON3brane}) then can be rewritten as
\be \label{EinsteinLambda.eq.3brane}
G_{\mu \nu }  + \Lambda g_{\mu \nu }  = 8\pi T_{\mu \nu }  + 64\pi ^2 \ell^2 S_{\mu \nu }  - E_{\mu \nu } \,.
\ee

To solve eq. (\ref{EinsteinLambda.eq.3brane}), we consider the Maxwell vector potential living on the brane to take the form
\be \label{Amu-ansatz}
A_\mu  {\rm{d}}x^\mu   =  - \frac{{Qr}}{{\rho ^2 }}\left( {{\rm{d}}u - \left( {a\Delta _x  - 2nx} \right){\rm{d}}\psi } \right)\,.
\ee 
Consequently, the corresponding field strength tensor related to the gauge field (\ref{Amu-ansatz}) reads
\be 
{\rm \bf F} = \frac{Q}{{\rho ^4 }}\left[ {\left( {r^2  - \left( {n + ax} \right)^2 } \right){\rm{d}}r \wedge \left( {{\rm{d}}u - P{\rm{d}}\psi } \right) - 2r\left(n+ax\right){\rm{d}}x \wedge \left( {\left( {r^2  + a^2  + n^2 } \right){\rm{d}}\psi  - a{\rm{d}}u} \right)} \right]
\ee
which is just (\ref{FmnQtrxphi}) after replacing ${\rm d}t \to {\rm d}u$ and ${\rm d}\phi \to {\rm d}\psi$. Clearly the energy-momentum tensor $T_{\mu\nu}$ associated to this field strength tensor is traceless, but the trace of $S_{\mu\nu}$ can be found to be
\be
S = \frac{{Q^4 }}{{64\pi ^2 \rho ^8 }}\,.
\ee 
The Hamiltonian constraint in this case achieved by taking the trace of eq. (\ref{EinsteinLambda.eq.3brane}) reads
\be \label{HamiltonRSLambda}
R + \kappa _5^4 S = 4\Lambda\,.
\ee  
Applying the last equation to the metric ansatz (\ref{metric.KerrSchildmNUTLambda}) gives us an equation,
\be\label{eq.H.chargedrotating}
\frac{{\partial ^2 H\left( {r,x} \right)}}{{\partial r^2 }} + \frac{4r}{{\rho ^2 }}\frac{{\partial H\left( {r,x} \right)}}{{\partial r}} + \frac{{2H\left( {r,x} \right)}}{{\rho ^2 }} =  - \frac{{\ell^2 Q^4 }}{{\rho ^8 }}
\ee 
whose solution can be written as
\be \label{HsolGEN}
H\left( {r,x} \right) = \frac{{2Mr - Q^2  - \beta  - \ell^2 Q^4 h\left( {r,x} \right)}}{{\rho ^2 }} \,,
\ee 
where
\be \label{hsolGEN}
h\left( {r,x} \right) = \frac{1}{{8\left( {ax + n} \right)^4 }}\left\{ {2 + \frac{{r^2 }}{{\rho ^2 }} + \frac{{3r}}{{\left( {ax + n} \right)}}\tan ^{ - 1} \left( {\frac{r}{{ax + n}}} \right)} \right\}\,.
\ee 
The solution (\ref{HsolGEN}) reduces to the one found in \cite{Aliev:2005bi} for the null NUT parameter $n$.  

To bring this solution into the form of Boyer-Lindquist type metric, we can employ the following transformation 
\be \label{BLtransf1}
{\rm d}u = {\rm d}t + \frac{{\left( {r^2  + a^2  + n^2 } \right)}}{{{{\Delta}_{r,Q}} }}{\rm d}r~~,~~{\rm d}\psi  = {\rm d}\phi  + \frac{a}{{{{\Delta}_{r,Q}} }}{\rm d}r\,,
\ee 
with
\be 
\Delta _{r,Q}  =  \Delta _{r,\Lambda}  + Q^2  + \ell^2 Q^4 h_0 \,,
\ee 
and $h_0 = h\left(x_0\right)$. The resulting metric after performing the above transformation reads
\[
{\rm d}s^2  =  - \frac{{\left( {{{\Delta}_{r,Q}}  + \delta  - a^2 \Xi \Delta _x } \right)}}{{\rho ^2 }}{\rm d}t^2   + \frac{{\left( {\Xi \Delta _x \left( {r^2  + a^2  + n^2 } \right)^2  - P^2 \left( {{{\Delta}_{r,Q}}  + \delta } \right)} \right)}}{{\rho ^2 }}{\rm d}\phi ^2  + \rho ^2 \frac{{dx^2 }}{{\Xi \Delta _x }}
\]
\be \label{BWBHfull1}
+ \rho ^2 \left( {1 - \frac{\delta }{{ \Delta _{r,Q} }}} \right)\frac{{dr^2 }}{{ \Delta _{r,Q} }} + \frac{{2\delta }}{{{{\Delta}_{r,Q}} }}\left( {P{\rm d}\phi  - {\rm d}t} \right){\rm d}r + \frac{{2\left( {P\left( {{{\Delta}_{r,Q}}  + \delta } \right) - a\Delta _x \Xi \left( {r^2  + a^2  + n^2 } \right)} \right)}}{{\rho ^2 }}{\rm d}t{\rm d}\phi \,,
\ee 
with $\delta  = \ell^2 Q^4 \left( {h - h_0 } \right)$. In the absence of NUT parameter $n$ and cosmological constant $\Lambda$, the metric above is just the charged and rotating spacetime on the brane reported in \cite{Aliev:2005bi}. 

An alternative coordinate transformation to get the Boyer-Lindquist type metric is 
\be \label{BLtransf2}
{\rm{d}}u = {\rm{d}}t + \frac{{\left( {r^2  + a^2  + n^2 } \right)\left( { \Delta _{r,Q}  - \delta } \right){\rm{d}}r}}{{ \Delta _{r,Q} \left( { \Delta _{r,Q}  + \delta } \right)}}~~,~~{\rm{d}}\psi  = {\rm{d}}\phi  + \frac{{a\left( { \Delta _{r,Q}  - \delta } \right){\rm{d}}r}}{{ \Delta _{r,Q} \left( { \Delta _{r,Q}  + \delta } \right)}} \,,
\ee
where the spacetime metric now reads
\[
{\rm d}s^2  =  - \frac{{\left( {{{\Delta}_{r,Q}}  + \delta  - a^2 \Xi \Delta _x } \right)}}{{\rho ^2 }}{\rm d}t^2   + \frac{{\left( {\Xi \Delta _x \left( {r^2  + a^2  + n^2 } \right)^2  - P^2 \left( {{{\Delta}_{r,Q}}  + \delta } \right)} \right)}}{{\rho ^2 }}{\rm d}\phi ^2  + \rho ^2 \frac{{dx^2 }}{{\Xi \Delta _x }}
\]
\be \label{BWBHfull2}
+ \rho ^2 \left( {1 - \frac{\delta }{{ \Delta _{r,Q} }}} \right)\frac{{dr^2 }}{{ \Delta _{r,Q} }} - \frac{{2\delta }}{{{{\Delta}_{r,Q}} }}\left( {P{\rm d}\phi  - {\rm d}t} \right){\rm d}r + \frac{{2\left( {P\left( {{{\Delta}_{r,Q}}  + \delta } \right) - a\Delta _x \Xi \left( {r^2  + a^2  + n^2 } \right)} \right)}}{{\rho ^2 }}{\rm d}t{\rm d}\phi \,.
\ee 
The last expression differs to (\ref{BWBHfull1}) in the signs of $\left( {t,r} \right)$ and $\left( {\phi,r} \right)$ components of the metric. Nevertheless, by doing the transformation $\phi\to -\phi$ and $t\to -t$, the result (\ref{BWBHfull2}) transforms to the one in (\ref{BWBHfull1}). This behavior is also the case for the charged and rotating black hole metric on the brane found by Aliev et al. \cite{Aliev:2005bi}. In the absence of NUT parameter $n$, the asymptotics of the spacetime metric above is de Sitter for $\Lambda >0$, and anti-de Sitter when $\Lambda <0$. 

Now let us discuss the accompanying vector field to this spacetime metric in Boyer-Lindquist type coordinate. The transformed vector field (\ref{Amu-ansatz}) using the transformation (\ref{BLtransf1}) reads
\be\label{AmuBL1-x}
A_\mu  {\rm{d}}x^\mu   =  - \frac{{Qr}}{{\rho ^2 }}\left( {{\rm{d}}t - \left( {a\Delta _x  - 2nx} \right){\rm{d}}\phi } \right) - \frac{{Qr}}{{\Delta _{r,Q} }}{\rm{d}}r\,.
\ee 
At first sight, this vector is not what we would expect for a charged rotating spacetime with timelike and axial Killing symmetries. Nevertheless, the gauge freedom allows one to shift the gauge field  
\be 
A_\mu  {\rm{d}}x^\mu   \to A_\mu  {\rm{d}}x^\mu   + A_r \left( r \right){\rm{d}}r
\ee
leaving the field strength tensor $F_{\mu\nu}$ remains unchanged. Therefore, we are permitted to get rid of the $A_r$ component in (\ref{AmuBL1-x}) to get a solution which reduces to the vector field of Kerr-Newman system if all of the bulk contributions such as tidal charge $\beta$ and AdS$_5$ radius $\ell$ are turned off, if any. In fact, we observe that the gauge field (\ref{AmuBL1-x}) does not contain any of these bulk parameters, which is understood since this field is a local property of the brane. Now, we can just express the vector solution companion to the line element (\ref{BWBHfull1}) to be
\be \label{AmuBL1}
A_\mu  {\rm{d}}x^\mu   =  - \frac{{Qr}}{{\rho ^2 }}\left( {{\rm{d}}t - \left( {a\Delta _x  - 2nx} \right){\rm{d}}\phi } \right)\,.
\ee 

So now we can verify that by taking the bulk parameters $\ell$ and $\beta$ vanish in the solution (\ref{BWBHfull1}), accompanied by the vector field (\ref{AmuBL1}), the fields solutions in Kerr-Newman-Taub-NUT-(A)dS system are recovered. It is interesting to find that the vector field (\ref{AmuBL1}) obeys the source free Maxwell equation $\nabla ^\mu  F_{\mu \nu }  = 0$, and obviously the Bianchi identity as well,
\be
\nabla _\kappa  F_{\mu \nu }  + \nabla _\nu  F_{\kappa \mu }  + \nabla _\mu  F_{\nu \kappa }  = 0\,,
\ee 
with the corresponding operator $\nabla_{\mu}$ above is performed in the background (\ref{BWBHfull1}). However, the transformation (\ref{BLtransf2}) brings the vector field (\ref{Amu-ansatz}) to the form
\be
A_\mu  {\rm{d}}x^\mu   =  - \frac{{Qr}}{{\rho ^2 }}\left( {{\rm{d}}t - \left( {a\Delta _x  - 2nx} \right){\rm{d}}\phi } \right) - \frac{{Qr\left( {\Delta _{r,Q}  - \delta } \right)}}{{\Delta _{r,Q} \left( {\Delta _{r,Q}  + \delta } \right)}}{\rm{d}}r
\ee 
where $\delta = \delta\left(r,x\right)$. The previous argument when we attempt to get rid of $A_r$ does not work when the extra term is a function of $x$, hence the transformation (\ref{BLtransf2}) is not the type that we would perform to get the solutions which reduce to the familiar form in Einstein-Maxwell system\footnote{See appendix \ref{app.KNnutAdS}.}.

\subsection{Static charged (A)dS braneworld black hole}\label{ssec.staticchargedAdS}

In this section, let us discuss how to achieve a static charged black hole in braneworld scenario with the AdS asymptotic, using the same approach employed previously. The obtained black hole solution would resemble the \RN-AdS black hole which has been studied extensively in aspects of AdS/CFT and holography \cite{Natsuume:2014sfa}. At first sight, the job seems to be trivial since we just need to set $n=0$ and $a=0$ in the solution (\ref{BWBHfull1}) and the corresponding gauge field (\ref{AmuBL1}) as well. However, these actions cannot be done directly in the corresponding spacetime, since it yields the singular result for the metric. The same concern also appeared in \cite{Aliev:2005bi}, where the authors had to expand their charged rotating solution in small rotation first, followed by imposing the static limit to match the static charged black hole solution in RS-II braneworld scenario reported in \cite{Chamblin}.

However, we could get the desired solution directly from the corresponding Hamiltonian constraint equation applied to the associated metric ansatz, rather than expanding the solution (\ref{BWBHfull1}) for small rotation and NUT parameter. The proper metric ansatz and gauge field are
\be\label{metric.AdSansatz} 
{\rm{d}}s^2  =  - \left( {1 - \frac{\Lambda_4 }{3}r^2 } \right){\rm{d}}u^2  + 2{\rm{d}}u{\rm{d}}r + r^2 \left( {\frac{{{\rm{d}}x^2 }}{{\Delta _x }} + \Delta _x {\rm{d}}\phi ^2 } \right) + H\left( {r,x} \right){\rm{d}}u^2 \,,
\ee
and
\be \label{Amu-AdS}
A_\mu  {\rm{d}}x^\mu   =  - \frac{Q}{r}{\rm{d}}u\,,
\ee 
respectively. The last two equations above are understood as the $n=0$ and $a=0$ cases of eqs. (\ref{metric.KerrSchildmNUTLambda}) and (\ref{AmuBL1}). Employing the Hamiltonian constrain (\ref{HamiltonRSLambda}) to this metric and gauge field, we can get an equation for $H\left(r,x\right)$,
\be\label{eq.H.chargedstatic}
\frac{{\partial ^2 H\left( {r,x} \right)}}{{\partial r^2 }} + \frac{4}{{r}}\frac{{\partial H\left( {r,x} \right)}}{{\partial r}} + \frac{{2H\left( {r,x} \right)}}{{r^2 }} =  - \frac{{\ell^2 Q^4 }}{{\rho ^8 }}\,.
\ee 
As one would have expected, eq. (\ref{eq.H.chargedstatic}) is the limits $a\to 0$ and $n\to 0$ of eq. (\ref{eq.H.chargedrotating}). The general solution to this equation is
\be\label{HsolstaticLambda}
H\left( {r,x} \right) = \frac{{F_1 \left( x \right)}}{r} + \frac{{F_2 \left( x \right)}}{{r^2 }} - \frac{{\ell^2 Q^4 }}{{20r^6 }}\,,
\ee 
with $F_1 \left( x \right)$ and $F_2 \left( x \right)$ are some $x$ dependent functions to be determined. In order to have agreement between the solution (\ref{metric.AdSansatz}) and the one presented in \cite{Chamblin} as $\Lambda_4$ is turned off, we have $F_1 \left( x \right) = 2m$ and $F_2 \left( x \right) = -\left(\beta+Q^2\right)$. 

Furthermore, to get the diagonal form of the metric we can perform the transformation  
\be \label{transf.ChargedAdS}
{\rm d} u\to {\rm d} t + \frac{r^2}{\Delta} {\rm d}r\,,
\ee 
where
\be \label{DeltaBWBHRNAdS}
\Delta  = r^2  - 2mr + \beta  + Q^2  - \frac{{\Lambda_4 r^4 }}{3} + \frac{{\ell^2 Q^4 }}{{20r^4 }}\,.
\ee 
This transformation yields the spacetime metric (\ref{metric.AdSansatz}) with the $H\left(r,x\right)$ function as given in (\ref{HsolstaticLambda}) can be written as
\be\label{metric.BWBHRNAdS} 
{\rm{d}}s^2  =  - F\left( r \right){\rm{d}}t^2  + F\left( r \right)^{ - 1} {\rm{d}}r^2  + r^2 {\rm{d}}\Omega^2 \,,
\ee
with 
\be 
F\left( r \right) = 1 - \frac{{2m}}{r} + \frac{{\beta  + Q^2 }}{{r^2 }} - \frac{{\Lambda_4 r^2 }}{3} + \frac{{\ell^2 Q^4 }}{{20r^6 }} \,,
\ee  
and ${\rm{d}}\Omega^2$ is the unit 2-sphere. 

It turns out that this solution had appeared in \cite{Sheykhi:2008et} where the topological charged braneworld black holes and some of their properties are discussed. The authors of \cite{Sheykhi:2008et} follow the diagonal metric ansatz proposed in \cite{Chamblin}, and solve the metric function using the corresponding Hamiltonian constraint equation (\ref{HamiltonRSLambda}). In this section we just showed that the same result can be achieved using the Kerr-Schild ansatz for the metric. Related to the vector field (\ref{Amu-AdS}), the transformation (\ref{transf.ChargedAdS}) yields this field can be expressed as
\be 
A_\mu {\rm d}x^\mu = - \frac{Q}{r} {\rm d}t\,,
\ee 
where the redundant $A_r \left(r\right)$ component has been removed using the gauge freedom argument. This gauge potential is exactly the vector fields of \RN-(A)dS of Einstein-Maxwell theory. In the vanishing of $\Lambda_4$, this metric reduces to that of Chamblin et al. \cite{Chamblin}, and turning off parameters form the bulk $\ell$ and $\beta$ simultaneously gives us the Reissner-Nordstrom-(A)dS spacetime. Note that the spacetime metric (\ref{metric.BWBHRNAdS}) can be asymptotically dS or AdS, depending on the value of $\Lambda_4$. 

\section{Geodesics around braneworld black holes with NUT parameter}\label{sec.geodesicsaroundblackhole}

\subsection{Hamilton-Jacobi equation}

Studies on timelike and null geodesics around rotating object play an important role in gravitational physics \cite{DeWitt:1973uma,Bardeen:1972fi,Aliev:2004ds,Chakraborty:2019rna,Mukherjee:2018dmm,Cebeci:2017xex,Siahaan:2019kbw,Jefremov:2016dpi}. It may give us predictions on reasonable orbit for some astrophysical objects or even light ring radius. Moreover, recently studies on these geodesics may provide us some estimates on black hole mergers \cite{Jai-akson:2017ldo,Siahaan:2019oik}. Particularly for circular geodesics, one can employ the approach by writing down the corresponding radial effective potential directly from the metric as performed in \cite{Siahaan:2019oik,Jai-akson:2017ldo}, or alternatively from the associated Hamilton-Jacobi equation for a test particle. Both approaches should lead to the same conclusion. 

Here we proceed to study the geodesics of null and timelike test particles in the background (\ref{metric.KNUT.brane}) by obtaining the Hamilton-Jacobi equation first. The corresponding horizon that corresponds to this spacetime has the radius
\be 
r_+   = M + \sqrt {M^2  + n^2  - \beta  - a^2 } \,,
\ee 
which is one of the zeros of $\Delta_r$ in the metric (\ref{metric.KNUT.brane}). The general form of Lagrangian for a test body can be read as
\be\label{Lang.test}
{\cal L} = \frac{1}{2} g_{\mu \nu } \dot x^\mu  \dot x^\nu  \,,
\ee 
where the overdot stands for the differentiation with respect to the affine parameter $\lambda$. As usual, the proper time $\tau$ is related to the affine parameter via $\tau = m\lambda$. Due to the metric convention in (\ref{metric.KNUT.brane}), our normalization condition is $g_{\mu \nu } \dot x^\mu  \dot x^\nu = \gamma$ where $\gamma = -1$ for timelike and $\gamma=0$ for null. From the Lagrangian (\ref{Lang.test}), the four momentum of test particle can be expressed as
\be 
p_\mu   = \frac{{\partial {\cal L}}}{{\partial \dot x^\mu  }}\,,
\ee 
where the relativistic relation $p^2 =- m^2$ of a particle with mass $m$ holds. On the other hand, the Hamilton-Jacobi equation can be written as \cite{DeWitt:1973uma}
\be \label{HJeq}
- \frac{{\partial {\cal A}}}{{\partial \lambda }} = \frac{1}{2}g^{\mu \nu } \partial _\mu  {\cal A}\partial _\nu  {\cal A}\,,
\ee 
where $\cal A$ is an action for the test particle satisfying
\be \label{pmuA}
p_\mu   = \partial _\mu  {\cal A}\,, 
\ee
and
\be \label{dtAm2}
\frac{{\partial {\cal A}}}{{\partial \tau }} = \frac{{m^2 }}{2}\,.
\ee 

Just like the ordinary Kerr-Taub-NUT solution, the spacetime (\ref{metric.KNUT.brane}) also possesses the timelike and axial Killing vectors. Consequently, there exist two constants of motion that belong to a test body in this spacetime, namely the energy $E=-p_t$ and angular momentum $L=p_\phi$. To proceed the Hamilton-Jacobi equation (\ref{HJeq}), we can employ the ansatz 
\be \label{HJanstz}
{\cal A} = \frac{{m^2 \tau }}{2} - Et + L\phi  + {\cal A}_r \left( r \right) + {\cal A}_x \left( x \right)\,,
\ee 
which agrees to (\ref{pmuA}) and (\ref{dtAm2}). Accordingly, we now have
\be 
E = \frac{{\Delta _r  - a\Delta _x }}{{\rho ^2 }}p^t  - \frac{{P\Delta _r  - a\left( {\rho ^2  + aP} \right)\Delta _x }}{{\rho ^2 }}p^\phi  \,,
\ee 
and
\be 
L = \frac{{P\Delta _r  - a\left( {\rho ^2  + aP} \right)\Delta _x }}{{\rho ^2 }}p^t  - \frac{{P^2 \Delta _r  - \left( {\rho ^2  + aP} \right)^2 \Delta _x }}{{\rho ^2 }}p^\phi  \,.
\ee
Solving the last two equations yields
\be 
p^t = \frac{{P\left( {L - PE} \right)}}{{\rho ^2 \Delta _x }} + \frac{{\left( {aP + \rho ^2 } \right)\left( {\left( {\rho ^2  + aP} \right)E - aL} \right)}}{{\rho ^2 \Delta _r }}\,,
\ee
and
\be 
p^\phi  = \frac{{L - PE}}{{\rho ^2 \Delta _x }} + \frac{{a\left( {\left( {\rho ^2  + aP} \right)E - aL} \right)}}{{\rho ^2 \Delta _r }}\,.
\ee

Explicitly, the Hamilton-Jacobi equation (\ref{HJeq}) can now be expressed as
\[
\Delta _r \left( {\frac{{\partial {\cal A}_r }}{{\partial r}}} \right)^2  + m^2 \left( {r^2  + n^2 } \right) + \left( {L - aE} \right)^2  - \frac{1}{{\Delta _r }}\left( {\left( {r^2  + n^2  + a^2 } \right)E - aL} \right)^2 
\]
\be 
=  - \Delta _x \left( {\frac{{\partial {\cal A}_x }}{{\partial x}}} \right)^2  + x^2 \left( {a^2 \left( {E^2  - m^2 } \right) - \frac{{L^2 }}{{\Delta _x }}} \right) + 2anx\left( {2E^2  - m^2 } \right) - \frac{{4nxE}}{{\Delta _x }}\left( {nxE + L} \right)\,.
\ee 
From the equation above, we can have
\be \label{eqCarterR}
\Delta _r \left( {\frac{{\partial {\cal A}_r }}{{\partial r}}} \right)^2  + m^2 \left( {r^2  + n^2 } \right) + \left( {L - aE} \right)^2  - \frac{1}{{\Delta _r }}\left( {\left( {r^2  + n^2  + a^2 } \right)E - aL} \right)^2  =  - K\,,
\ee 
and
\be \label{eqCarterX}
\Delta _x \left( {\frac{{\partial {\cal A}_x }}{{\partial x}}} \right)^2  + x^2 \left( {a^2 \left( {m^2  - E^2 } \right) + \frac{{L^2 }}{{\Delta _x }}} \right) + 2anx\left( {m^2  - 2E^2 } \right) + \frac{{4nxE}}{{\Delta _x }}\left( {nxE + L} \right) = K\,,
\ee 
where $K$ is known as Carter constant. Since $p_\mu = \partial_\mu {\cal A}$, from (\ref{eqCarterR}) and (\ref{eqCarterX}) we can write
\be p_x  =  \pm \rho ^{ - 2} \sqrt {{\cal X}\left( x \right)}\,, \ee 
and  
\be p_r  =  \pm \rho ^{ - 2} \sqrt {{\cal R}\left( r \right)}\,, \ee  
with
\be \label{X-eq}
{\cal X}\left( x \right) = \Delta _x K - x^2 \left( {a^2 \left( {m^2  - E^2 } \right)\Delta _x  + L^2 } \right) + 2anx\Delta _x \left( {m^2  - 2E^2 } \right) + 4nxE\left( {nxE + L} \right)\,,
\ee 
and
\be \label{R-eq}
{\cal R}\left( r \right) = \left( {\left( {r^2  + n^2  + a^2 } \right)E - aL} \right)^2  - \Delta _r \left( {K + m^2 \left( {r^2  + n^2 } \right) + \left( {L - aE} \right)^2 } \right)\,.
\ee
In the absence of NUT parameter $n$ and tidal charge $\beta$, the last two equations reduce to those of Kerr spacetime \cite{DeWitt:1973uma}. In the next sections, we will use theses results above to investigate the timelike and null circular geodesics, for equatorial and non-equatorial planes.

\subsection{Timelike geodesic}

Motion on the plane for a fixed $x=x_0$ must obey \cite{Jefremov:2016dpi}
\be \label{eqX}
{\cal X}\left( x_0 \right) = 0~~{\rm and}~~\left. {\frac{{d{\cal X}\left( x \right)}}{{dx}}} \right|_{x = x_0 } = 0\,,
\ee 
where ${\cal X}\left(x\right)$ is given in (\ref{X-eq}). The two equations in (\ref{eqX}) are related to the velocity and acceleration in $x$ direction, namely $\dot x$ and $\ddot x$ respectively. Obviously, the first equation in (\ref{eqX}) is satisfied in equatorial plane $x=0$ since the Carter constant vanish in this consideration as dictated by eq. (\ref{eqCarterX}).  However, the second equation in (\ref{eqX}) which can be expressed as
\be \label{dXdx}
\left. {\frac{{d{\cal X}\left( x \right)}}{{dx}}} \right|_{x = 0 } = 2n\left( {a\left( {m^2  - 2E^2 } \right) + 2EL} \right)\,,
\ee 
which vanishes in the absence of the NUT parameter $n$. This is exactly the case of the neutral rotating braneworld black hole with a tidal charge $\beta$ studied in \cite{Aliev:2005bi}, i.e. the equatorial timelike geodesics are guaranteed. For non-zero $n$, the last equation constrains the properties of test particles and spacetime to yield the equatorial radial geodesics exist. This in turn becomes a stringent restriction for the existence of equatorial circular orbits. 

Now let us define some effective potentials related to the motion in $r$ and $x$ directions, namely \cite{Bardeen:1972fi}
\be \label{drdtVrVx}
\rho ^4 \left( {\frac{{dr}}{{d\tau }}} \right)^2  = V_{r,{\rm{eff}}}~~ {\rm{  and  }}~~\rho ^4 \left( {\frac{{dx}}{{d\tau }}} \right)^2  = V_{x,{\rm{eff}}} \,,
\ee 
from which we can write
\be \label{Vr-eff}
V_{r,{\rm{eff}}} =\left( {\left( {r^2  + n^2  + a^2 } \right){\epsilon} - ah} \right)^2  - \Delta _r \left( {k +\left( {r^2  + n^2 } \right) + \left( {h - a\epsilon} \right)^2 } \right)\,,
\ee 
and
\be \label{Vx-eff}
V_{x,{\rm{eff}}} = \Delta _x k - x^2 \left( {a^2 \left( {1  - \epsilon^2 } \right)\Delta _x  + h^2 } \right) + 2anx\Delta _x \left( {1  - 2\epsilon^2 } \right) + 4nx\epsilon\left( {nx\epsilon + h} \right)\,.
\ee 
In equations above we have used $\epsilon \equiv Em^{-1}$, $h \equiv L m^{-1}$, and $k\equiv Km^{-1}$. To obtain the innermost stable circular orbit (ISCO) radius, the set of conditions
\be \label{eqVrTL}
V_{r,{\rm{eff}}}  = 0~~,~~\frac{{dV_{r,{\rm{eff}}} }}{{dr}} = 0~~,~~\frac{{d^2 V_{r,{\rm{eff}}} }}{{d^2 r}} = 0\,,
\ee 
and
\be \label{eqVxTL}
\left. {V_{x,{\rm{eff}}} } \right|_{x = 0}  = 0~~,~~\left. {\frac{{dV_{x,{\rm{eff}}} }}{{dx}}} \right|_{x = 0}  = 0\,,
\ee 
must be satisfied simultaneously. Equations (\ref{eqVrTL}) are related to the motion in radial direction, and equations (\ref{eqVxTL}) to keep that motion on the plane with a particular fixed $x$. It is obvious that $V_{r,{\rm{eff}}}$ is independent of $x$ coordinate, as $V_{x,{\rm{eff}}}$ does not depend on $r$. The first equation in (\ref{eqVxTL}) is satisfied by effective potential (\ref{Vx-eff}). As indicated before, normally equations (\ref{eqVxTL}) do not appear when one studies rotating black hole geometries without NUT parameter since the nature of that spacetime that guarantees the motion on equatorial plane. However, the second one in (\ref{eqVxTL}) is satisfied for zero NUT parameter, or when
\be \label{constrain-dVx-TL}
a\left( {1 - 2\epsilon^2 } \right) + 2\epsilon h =0
\ee 
for non zero NUT parameter, which can be understood from eq. (\ref{dXdx}). The last equation can be considered as a new additional constraint to the set equations in (\ref{eqVrTL}) to find the ISCO radius on equatorial plane. Nevertheless, from (\ref{constrain-dVx-TL}) and the set of equations in (\ref{eqVrTL}), one cannot discover a solution to ISCO radius supported by real valued NUT parameter. Consequently, we can infer that equatorial timelike circular geodesics cannot occur in the spacetime (\ref{metric.KNUT.brane}), unlike in the case of null NUT parameter \cite{Aliev:2005bi}. 

\begin{figure}
	\begin{center}
		\includegraphics*[scale=0.4]{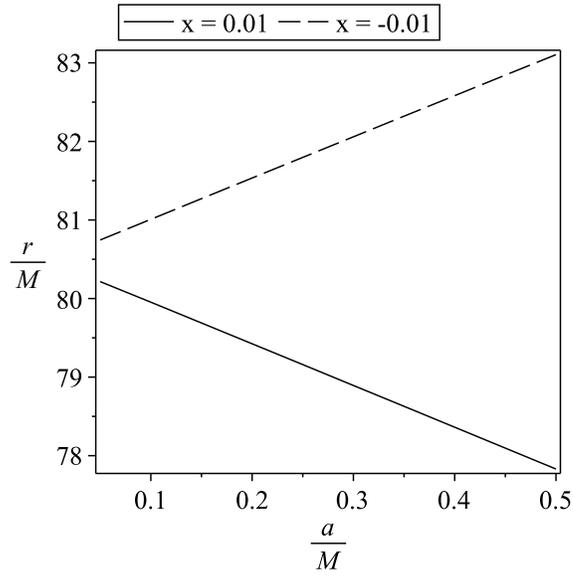}
	\end{center}
\caption{ISCO radius on the $x=0.01$ and $x=-0.01$ planes, and the numerical value for the NUT parameter $n=0.1$.} \label{Fig.rn01xpm001}
\end{figure}

\begin{figure}
	\begin{center}
		\includegraphics*[scale=0.4]{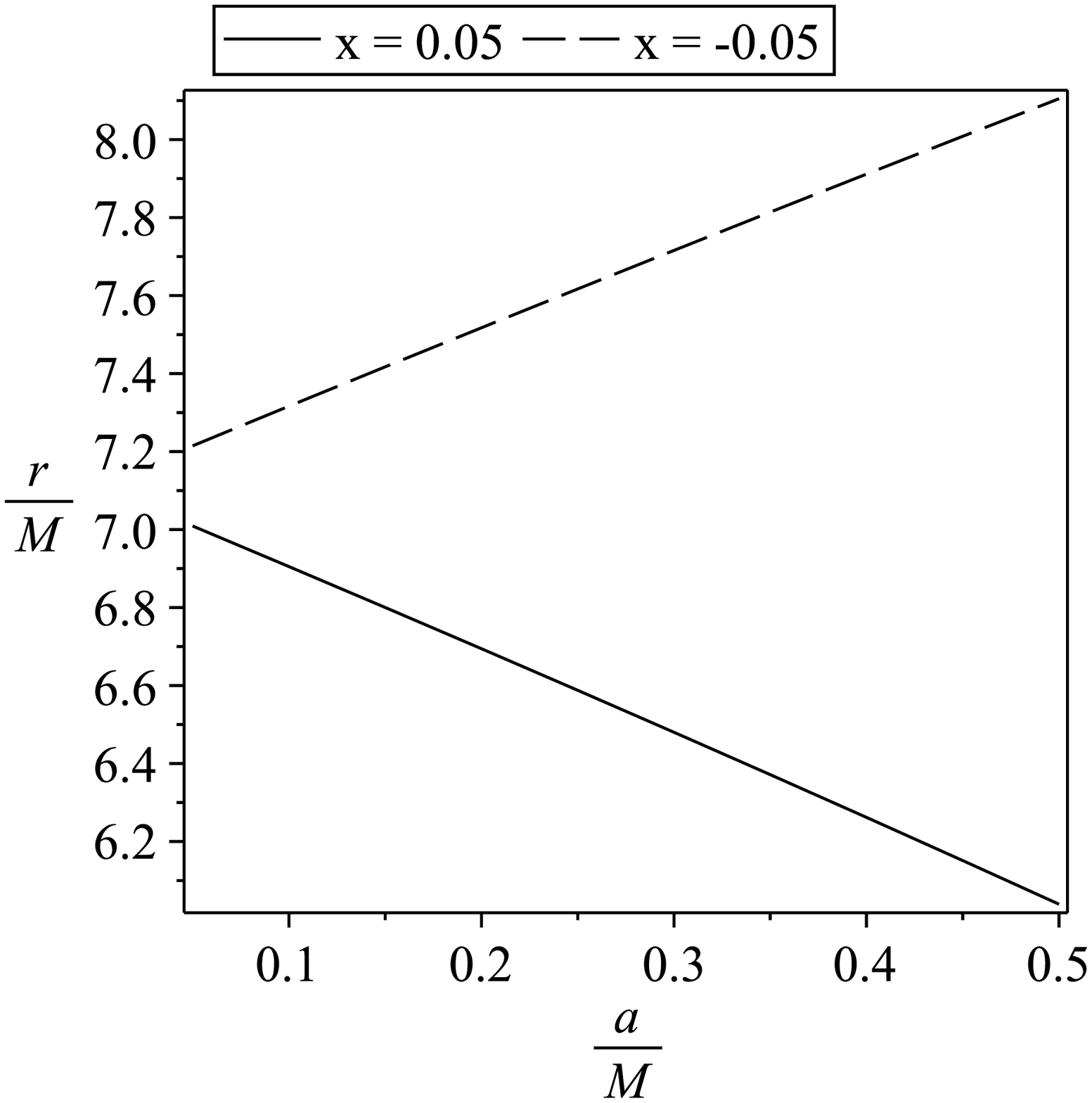}
	\end{center}
\caption{ISCO radius on the $x=0.05$ and $x=-0.05$ planes, and the numerical value for the NUT parameter $n=0.1$.} \label{Fig.rn01xpm005}
\end{figure}

\begin{figure}
	\begin{center}
		\includegraphics*[scale=0.4]{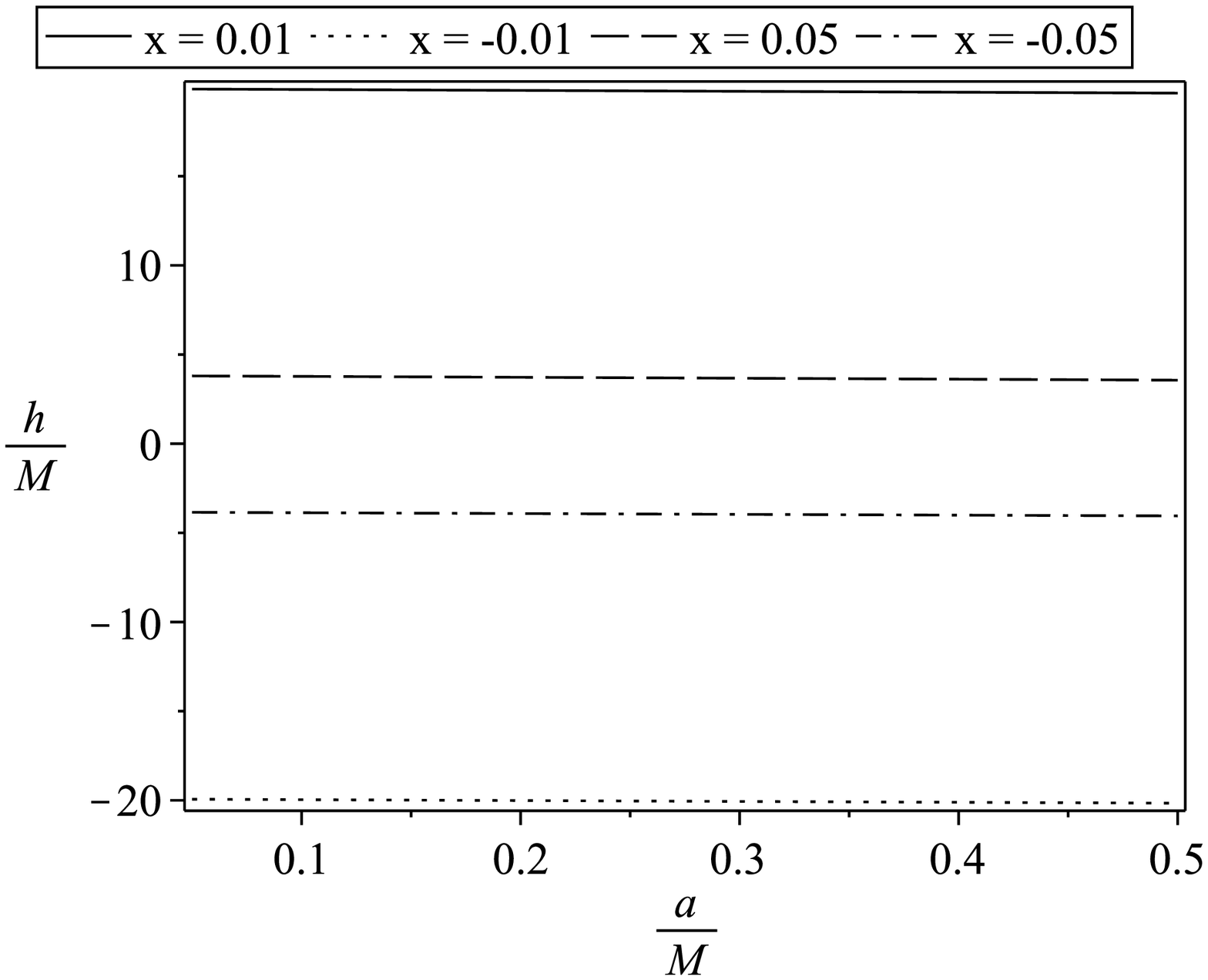}
	\end{center}
\caption{The angular momentum $h$ for each cases plotted in figures \ref{Fig.rn01xpm001} and \ref{Fig.rn01xpm005}.} \label{Fig.hn01xpm001pm005}
\end{figure}

However the timelike circular geodesics could lie on some non-equatorial planes. Investigating this analytically is obviously troublesome, hence we would pursue this job numerically  as presented in figs. \ref{Fig.rn01xpm001} and \ref{Fig.rn01xpm005}. From these plots we learn that such geodesics exist, despite the incorporating parameters such as the NUT parameter $n$ and tidal charge $\beta$ are severely constrained. All ISCO radii presented in figs. \ref{Fig.rn01xpm001} and \ref{Fig.rn01xpm005} are larger than the corresponding event horizon, and the associated tidal charge parameter $\beta$ are all negative. Note that the negative tidal charge is considered to be the physically more natural one \cite{Dadhich}. Interestingly, from fig. \ref{Fig.hn01xpm001pm005} we find that circular geodesics for $x=0.01$ and $x=0.05$ are in prograde motion, while for $x=-0.01$ and $x=-0.05$ are in retrograde motion. 

\subsection{Null geodesic}

In this section we discuss the possibility to observe light ring around the rotating braneworld spacetime with NUT parameter. The corresponding effective potential in null case can be achieved by taking the massless limit of equations (\ref{X-eq}) and (\ref{R-eq}), \be \label{drdtVrVx-null}
\rho ^4 \left( {\frac{{dr}}{{d\lambda }}} \right)^2  = V_{r,{\rm{eff}}}~~ {\rm{  and  }}~~\rho ^4 \left( {\frac{{dx}}{{d\lambda }}} \right)^2  = V_{x,{\rm{eff}}} \,,
\ee 
from which we can write
\be \label{Vr-eff-null}
V_{r,{\rm{eff}}} =\left( {\left( {r^2  + n^2  + a^2 } \right){\epsilon} - ah} \right)^2  - \Delta _r \left( {k + \left( {h - a\epsilon} \right)^2 } \right)\,,
\ee 
and
\be \label{Vx-eff-null}
V_{x,{\rm{eff}}} = \Delta _x k + x^2 \left( {a^2 \epsilon^2 \Delta _x  - h^2 } \right) +4nx\epsilon \left(\epsilon\left(nx-a\Delta_x\right)+ h\right)\,,
\ee 
where again we have used the notations $\epsilon \equiv Em^{-1}$, $h \equiv L m^{-1}$, and $k\equiv Km^{-1}$. Nevertheless, the set of equations 
\be \label{VrVxeqtnNULL}
V_{x,{\rm{eff}}}  = 0~,~\frac{{dV_{x,{\rm{eff}}} }}{{dx}} = 0~,~V_{r,{\rm{eff}}}  = 0~,~\frac{{dV_{r,{\rm{eff}}} }}{{dr}} = 0\,,
\ee 
for the effective potentials given in (\ref{Vr-eff-null}) and (\ref{Vx-eff-null}) do not have solutions for radius $r$ for the case $x=0$, which implies that light ring does not exist on equatorial plane in rotating braneworld spacetime holes with NUT parameter.  

\begin{figure}
	\begin{center}
		\includegraphics*[scale=0.4]{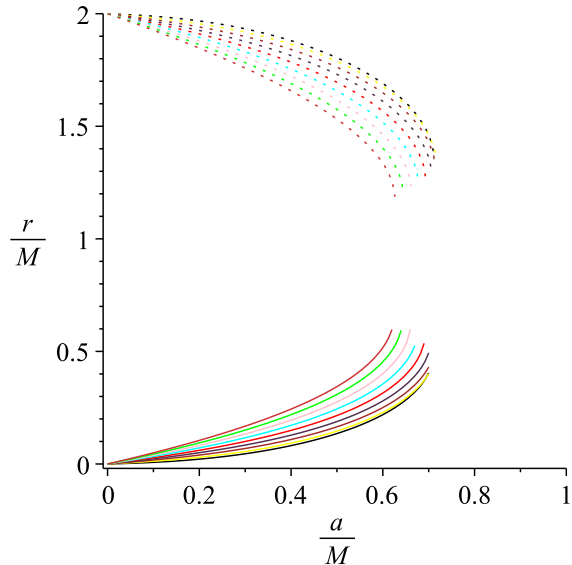}
	\end{center}
	\caption{The case of $x=0.1,0.2,0.3,0.4,0.5,0.6,0.7,0.8,0.9$ are represented by the color black, yellow, brown, violet, red, cyan, pink, green, orange, respectively. The solid lines are the circular radii, and the dashed ones are the corresponding horizons. The associated $\beta$ and NUT parameters for each lines in these plots are real valued.} \label{Fig.rnullplus}
\end{figure}

Now, as we did for timelike case previously, let us investigate the possibility for light rights to exist in non-equatorial plane. To do so, we must deal with the set of equations (\ref{VrVxeqtnNULL}) for $x \neq 0$, where some numerical evaluations are given in fig. \ref{Fig.rnullplus}. However, these results do not support the existence of null circular geodesics on the evaluated non-equatorial planes, since all of the achieved radii in fig. \ref{Fig.rnullplus} are inside the corresponding horizons. Furthermore, from the variety of evaluated non-zero $x$ under consideration, we can claim that it is unlikely for a rotating braneworld spacetime equipped with NUT parameter to accommodate light rings on non-equatorial planes.

\section{Conclusion and Discussion}

In this work, we have generalized the braneworld spacetime solution in RS-II model obtained by Aliev et al. \cite{Aliev:2005bi} and Neves et al. \cite{Neves:2012it}. The effective Einstein equation on the brane is the one discovered by Shiromizu et al. \cite{Shiromizu}, and the method to get the solutions presented in this paper closely follows the approach introduced in \cite{Aliev:2005bi}. To obtain the solutions presented in section \ref{sec.sol3brane}, we have made several assumptions. First we assume the existence of bulk contribution  through $E_{\mu\nu}$ that yields the effective Einstein equations on the brane to be closed. Indeed, it is remarkable if we can write down a bulk metric that corresponds to this $E_{\mu\nu}$ and the brane metric $g_{\mu\nu}$, but we do not aim to pursue this question in our present work. This concern also arises in the preceding works \cite{Chamblin,Aliev:2005bi}. Secondly, for the non-vacuum brane, we assume that the Maxwell field to be localized on the brane. This field will give rise to the electric charge of a black hole. Finally, as proposed in \cite{Aliev:2005bi}, we also assume that the brane allows the Kerr-Schild form of its spacetime metric. Using these asumptions, we manage to obtain several solutions in section \ref{sec.sol3brane}, based the effective Einstein equation on a brane by Shiromizu et al. \cite{Shiromizu}. 

In section \ref{sec.geodesicsaroundblackhole}, we study the circular geodesics for timelike and null test objects in the rotating braneworld spacetime with tidal charge and NUT parameter. We find that both timelike and null equatorial circular geodesics cannot occur in this background. However, numerical results presented in section \ref{sec.geodesicsaroundblackhole} support the existence of non-equatorial timelike circular geodesics, but not for the null case. Furthermore, the non-equatorial timelike circular geodesics in by the neutral rotating braneworld spacetime with NUT parameter can exist provided that the tidal charge is negative. Of course this result has no analog in Kerr-Newman-Taub-NUT spacetime case since it will require the charge $Q$ of the Kerr-Newman-Taub-NUT object to be purely imaginary.

As we have mentioned previously that the exact bulk metric which correspond to the solutions reported in this work has not been discovered, as also to the braneworld black holes solutions discovered in \cite{Dadhich,Chamblin,Aliev:2005bi,Neves:2012it}. It is going to be challenging to discover such bulk metric and to see how the localization for the black holes and incorporating Maxwell fields at works. In particular, here we also reproduce the static charged AdS braneworld black hole in RS-II model, which had been discussed in \cite{Sheykhi:2008et}. This type of black hole may have some interesting further studies, related to its thermodynamics \cite{Hawking:1982dh} and holographic uses \cite{Natsuume:2014sfa}. We address these projects in our future works.

\section*{Acknowledgement}

I thank Frisca G. D. Sinaga for her support and encouragement.

\appendix
\section{Kerr-Newman-Taub-NUT-(A)dS solution}\label{app.KNnutAdS}

The Kerr-Newman-Taub-NUT-(A)dS solution solves the Einstein equations with cosmological constant which is eq. (\ref{eq.EinsteinON3brane}) with vanishing $\kappa _5$ and $E_{\mu\nu}$. The line element can take the form \cite{Griffiths:2009dfa}
\be\label{metric.KNnutAdS}
{\rm{d}}s^2  =  - \frac{\cal Q}{{\rho ^2 }}\left[ {{\rm{d}}t - \left( {a\Delta _x  - 2nx} \right){\rm{d}}\phi } \right]^2  + \rho ^2 \left[ {\frac{{{\rm{d}}r^2 }}{\cal Q} + \frac{{{\rm{d}}x^2 }}{{{\cal P}\Delta }}} \right] + \frac{{{\cal P}\Delta _x }}{{\rho ^2 }}\left[ {a{\rm{d}}t - \left( {r^2  + a^2  + n^2 } \right){\rm{d}}\phi } \right]^2 \,,
\ee
with
\be 
{\cal Q} = r^2  - 2mr + a^2  + Q^2  - n^2  - \Lambda _4 \left[ {\left( {a^2  - n^2 } \right)n^2  + r^2 \left( {\frac{{a^2 }}{3} + 2n^2 } \right) + \frac{{r^4 }}{3}} \right]\,,
\ee 
and
\be
{\cal P} = 1 + \frac{4}{3}\Lambda alx + \frac{1}{3}\Lambda a^2 x^2 \,.
\ee 
The corresponding vector field is
\be 
A_\mu  {\rm{d}}x^\mu   = \frac{{Qr}}{{\rho ^2 }}\left[ {{\rm{d}}t - \left( {a\Delta _x  - 2nx} \right){\rm{d}}\phi } \right]\,.
\ee
Setting $n=0$ and $\Lambda=0$ in line element (\ref{metric.KNnutAdS}) yields the Kerr-Newman metric describing rotating mass with electric charge $Q$ in Einstein-Maxwell theory. The metric (\ref{metric.KNnutAdS}) obeys the Einstein equations
\be 
G_{\mu \nu }  + \Lambda _4 g_{\mu \nu }  = 8\pi T_{\mu \nu } \,,
\ee 
where the energy-momentum tensor is given in eq. (\ref{Tmn}).


\begin{thebibliography}{99}
	
\bibitem{Kiritsis:1997hj}
E.~Kiritsis,
\textit{Introduction to superstring theory},
[arXiv:hep-th/9709062 [hep-th]].

\bibitem{Townsend:1996xj}
P.~Townsend,
\textit{Four lectures on M theory},
[arXiv:hep-th/9612121 [hep-th]].

\bibitem{Becker}
K.~Becker, M.~Becker and J.~H.~Schwarz, \textit{String Theory and M-Theory: A Modern Introduction}, (Cambridge University Press, Cambridge, 2007).

\bibitem{Kaluza:1921tu}
T.~Kaluza,
Int. J. Mod. Phys. D \textbf{27}, no.14, 1870001 (2018)

\bibitem{Klein:1926tv}
O.~Klein,
Z. Phys. \textbf{37} (1926), 895-906

\bibitem{Rubakov:1983bb}
V.~Rubakov and M.~Shaposhnikov,
Phys. Lett. B \textbf{125}, 136-138 (1983)

\bibitem{ArkaniHamed:1998nn}
N.~Arkani-Hamed, S.~Dimopoulos and G.~R.~Dvali, Phys. Rev. D \textbf{59} (1999), 086004.

\bibitem{ArkaniHamed:1998rs}
N.~Arkani-Hamed, S.~Dimopoulos and G.~Dvali,
Phys. Lett. B \textbf{429}, 263-272 (1998)

\bibitem{Randall:1999ee}
L.~Randall and R.~Sundrum,
Phys. Rev. Lett. \textbf{83}, 3370-3373 (1999)

\bibitem{Randall:1999vf}
L.~Randall and R.~Sundrum,
Phys. Rev. Lett. \textbf{83}, 4690-4693 (1999)

\bibitem{Dvali:2000hr}
G.~Dvali, G.~Gabadadze and M.~Porrati,
Phys. Lett. B \textbf{485}, 208-214 (2000)

\bibitem{Karch}
A.~Karch and L.~Randall, J. High Energy Phys. \textbf{05} (2001), 008.

\bibitem{Maartens:2010ar}
R.~Maartens and K.~Koyama,
Living Rev. Rel. \textbf{13}, 5 (2010)

\bibitem{Tanahashi:2011xx}
N.~Tanahashi and T.~Tanaka,
Prog. Theor. Phys. Suppl. \textbf{189} (2011), 227-268

\bibitem{Shiromizu}
T.~Shiromizu, K.~I.~Maeda and M.~Sasaki, Phys. Rev. D \textbf{62} (2000), 
024012.

\bibitem{Dadhich} N. Dadhich, R. Maartens, P. Papadopoulos and V. Rezania,
Phys. Lett. B {\bf 487}, 1 (2000) 


\bibitem{Chamblin}
A.~Chamblin, H.~S.~Reall, H.~A.~Shinkai and T.~Shiromizu, Phys. Rev. D \textbf{63} (2001), 064015

\bibitem{Aliev:2005bi}
A.~N.~Aliev and A.~E.~Gumrukcuoglu,
Phys.\ Rev.\ D {\bf 71} (2005) 104027

\bibitem{Neves:2012it}
J.~C.~S.~Neves and C.~Molina,
Phys.\ Rev.\ D {\bf 86} (2012) 124047

\bibitem{Sheykhi:2008et}
A.~Sheykhi and B.~Wang,
Mod. Phys. Lett. A \textbf{24}, 2531-2538 (2009)

\bibitem{Kanti:2018ozd}
P.~Kanti, T.~Nakas and N.~Pappas,
Phys.\ Rev.\ D {\bf 98} (2018) no.6,  064025

\bibitem{Nakas:2019rod}
T.~Nakas, N.~Pappas and P.~Kanti,
Phys.\ Rev.\ D {\bf 99} (2019) no.12,  124040

\bibitem{Nakas:2020crd}
T.~Nakas, P.~Kanti and N.~Pappas,
Phys.\ Rev.\ D {\bf 101} (2020) no.8,  084056

\bibitem{Visinelli:2017bny}
L.~Visinelli, N.~Bolis and S.~Vagnozzi,
Phys. Rev. D \textbf{97}, no.6, 064039 (2018)

\bibitem{BinNun:2009jr}
A.~Y.~Bin-Nun,
Phys. Rev. D \textbf{81}, 123011 (2010)

\bibitem{Whisker:2004gq}
R.~Whisker,
Phys. Rev. D \textbf{71}, 064004 (2005)

\bibitem{Eiroa:2012fb}
E.~F.~Eiroa and C.~M.~Sendra,
Phys. Rev. D \textbf{86}, 083009 (2012)

\bibitem{Abdujabbarov:2017pfw}
A.~Abdujabbarov, B.~Ahmedov, N.~Dadhich and F.~Atamurotov,
Phys. Rev. D \textbf{96}, no.8, 084017 (2017)

\bibitem{Eiroa:2004gh}
E.~F.~Eiroa,
Phys. Rev. D \textbf{71}, 083010 (2005)

\bibitem{Vagnozzi:2019apd}
S.~Vagnozzi and L.~Visinelli,
Phys. Rev. D \textbf{100}, no.2, 024020 (2019)

\bibitem{Fitzpatrick:2006cd}
A.~Fitzpatrick, L.~Randall and T.~Wiseman,
JHEP \textbf{11}, 033 (2006)

\bibitem{Tanaka:2002rb}
T.~Tanaka,
Prog. Theor. Phys. Suppl. \textbf{148}, 307-316 (2003)

\bibitem{Emparan:2002px}
R.~Emparan, A.~Fabbri and N.~Kaloper,
JHEP \textbf{08}, 043 (2002)

\bibitem{Kanti:2001cj}
P.~Kanti and K.~Tamvakis,
Phys. Rev. D \textbf{65}, 084010 (2002)

\bibitem{Kanti:2003uv}
P.~Kanti, I.~Olasagasti and K.~Tamvakis,
Phys. Rev. D \textbf{68}, 124001 (2003)

\bibitem{Kanti:2013lca}
P.~Kanti, N.~Pappas and K.~Zuleta,
Class. Quant. Grav. \textbf{30}, 235017 (2013)

\bibitem{Kanti:2015poa}
P.~Kanti, N.~Pappas and T.~Pappas,
Class. Quant. Grav. \textbf{33}, no.1, 015003 (2016)

\bibitem{Fichet:2019owx}
S.~Fichet,
JHEP {\bf 2004} (2020) 016

\bibitem{Wang:2016nqi}
D.~Wang and M.~W.~Choptuik,
Phys. Rev. Lett. \textbf{117}, no.1, 011102 (2016)


\bibitem{Plebanski:1976gy}
J.~Plebanski and M.~Demianski,
Annals Phys. \textbf{98}, 98-127 (1976)

\bibitem{Griffiths:2009dfa}
J.~B.~Griffiths and J.~Podolsky,
\textit{Exact Space-Times in Einstein's General Relativity}, (Cambridge University Press, Cambridge, 2009)

\bibitem{Houri:2019lnu}
T.~Houri, N.~Tanahashi and Y.~Yasui,
Class. Quant. Grav. \textbf{37}, no.1, 015011 (2020)

\bibitem{Kolar:2019gzy}
I.~Kolar and P.~Krtous,
Phys. Rev. D \textbf{100}, no.6, 064014 (2019)

\bibitem{Chakraborty:2019rna}
C.~Chakraborty and S.~Bhattacharyya,
JCAP \textbf{05}, 034 (2019)

\bibitem{Long:2018tij}
F.~Long, S.~Chen, J.~Wang and J.~Jing,
Eur. Phys. J. C \textbf{79}, no.6, 466 (2019)

\bibitem{Frolov:2018eza}
V.~P.~Frolov and P.~Krtous,
Phys. Rev. D \textbf{99}, no.4, 044044 (2019)

\bibitem{Sadeghian:2018bli}
S.~Sadeghian,
Phys. Rev. D \textbf{98}, no.8, 084031 (2018)

\bibitem{Mukherjee:2018dmm}
S.~Mukherjee, S.~Chakraborty and N.~Dadhich,
Eur. Phys. J. C \textbf{79}, no.2, 161 (2019)

\bibitem{Krtous:2018bvk}
P.~Krtouš, V.~P.~Frolov and D.~Kubizňák,
Nucl. Phys. B \textbf{934}, 7-38 (2018)

\bibitem{Duztas:2017lxk}
K.~Düztaş,
Class. Quant. Grav. \textbf{35}, no.4, 045008 (2018)

\bibitem{Paganini:2017qfo}
C.~F.~Paganini and M.~A.~Oancea,
Class. Quant. Grav. \textbf{35}, no.6, 067001 (2018)

\bibitem{Cebeci:2017xex}
H.~Cebeci, N.~Özdemir and S.~Şentorun,
Gen. Rel. Grav. \textbf{51}, no.7, 85 (2019)

\bibitem{Cebeci:2015fie}
H.~Cebeci, N.~Ozdemir and S.~Sentorun,
Phys. Rev. D \textbf{93}, no.10, 104031 (2016)

\bibitem{Siahaan:2019kbw}
H.~M.~Siahaan,
[arXiv:1905.02622 [gr-qc]].


\bibitem{Sakti:2020jpo}
M.~Sakti, A.~Ghezelbash, A.~Suroso and F.~Zen,
Nucl. Phys. B \textbf{953}, 114970 (2020)

\bibitem{Sakti:2019zix}
M.~Sakti, A.~Ghezelbash, A.~Suroso and F.~Zen,
Gen. Rel. Grav. \textbf{51}, no.11, 151 (2019)

\bibitem{Sakti:2019krw}
M.~Sakti, A.~Suroso and F.~Zen,
Annals Phys. \textbf{413}, 168062 (2020)

\bibitem{Sakti:2017pmt}
M.~F.~A.~R.~Sakti, A.~Suroso and F.~P.~Zen,
Int. J. Mod. Phys. D \textbf{27}, no.12, 1850109 (2018)

\bibitem{Jefremov:2016dpi}
P.~Jefremov and V.~Perlick,
Class. Quant. Grav. \textbf{33}, no.24, 245014 (2016)

\bibitem{Aliev:2004ds}
A.~Aliev and A.~Gumrukcuoglu,
Class. Quant. Grav. \textbf{21}, 5081-5096 (2004)

\bibitem{Natsuume:2014sfa}
M.~Natsuume,
Lect. Notes Phys. \textbf{903}, pp.1-294 (2015)


\bibitem{DeWitt:1973uma}
C.~DeWitt and B.~S.~DeWitt,
``Proceedings, Ecole d'Eté de Physique Théorique: Les Astres Occlus : Les Houches, France, August, 1972,''

\bibitem{Bardeen:1972fi}
J.~M.~Bardeen, W.~H.~Press and S.~A.~Teukolsky,
Astrophys.\ J.\  \textbf{178}, 347 (1972)

\bibitem{Siahaan:2019oik}
H.~M.~Siahaan,
Phys. Rev. D \textbf{101}, no.6, 064036 (2020)

\bibitem{Jai-akson:2017ldo}
P.~Jai-akson, A.~Chatrabhuti, O.~Evnin and L.~Lehner,
Phys. Rev. D \textbf{96}, no.4, 044031 (2017)

\bibitem{Hawking:1982dh}
S.~Hawking and D.~N.~Page,
Commun. Math. Phys. \textbf{87}, 577 (1983)


\end{thebibliography}
\end{document}